\documentclass[12pt]{article}
\usepackage{a4,times,amssymb,epsf}

\newcommand{\Ss}[1]{{\sf\small #1}}

\begin{document}
\vspace{1.3in}
\begin{center}
{\Large {\bf Distributed Computation, the Twisted Isomorphism, and 
Auto-Poiesis}}\\
\vspace{0.2in}
{\large Michael Manthey}\\
{\small Computer Science Department\\Aalborg University\\Fr. Bajersvej 7E\\9200 
Aalborg, Denmark\\manthey@cs.auc.dk}\\
\end{center}

{\bf Abstract}
\footnote{Invited paper, CASYS'97 First International Conference on Computing 
Anticipatory Systems, Liege (Belgium), August 11-15, 1997. D. Dubois, Ed.
 \copyright \today.}

This paper presents a synchronization-based, multi-process computational model 
of anticipatory systems called the Phase Web. It describes a self-organizing 
paradigm that explicitly recognizes and exploits the existence of a boundary 
between inside and outside, accepts and exploits intentionality, and uses 
explicit self-reference to describe eg. auto-poiesis. The model explicitly 
connects  computation to a discrete Clifford algebraic formalization that is in 
turn extended into homology and co-homology, wherein the recursive nature of 
objects and boundaries becomes apparent and itself subject to hierarchical 
recursion. {\em Topsy}, a computer program embodying the Phase Web, is 
currently 
being readied for release.


\section*{Introduction}

Anticipatory systems (Rosen, 1985) display a number of properties that, 
together, differentiate them strongly from other kinds of systems:

\begin{itemize}
\item They possess {\em parts} that interact {\em locally} to form a coherently 
behaving {\em whole}.

\item The way in which these parts interact differ widely from system to system 
in detail, yet wholes with very different parts seem nevertheless to resemble 
each other {\em qua} their very wholeness.

\item It is impossible to ignore the fact that such systems are {\em situated} 
in a surrounding environment. Indeed, their interaction with their environment 
is so integral to what they are and do makes their very situatedness a defining 
characteristic.

\item A critical behavior shared by these wholes is the ability to {\em 
anticipate} changes in their surrounding environment and react in a way that 
(hopefully) ensures their continuing existence, ie. {\em auto-poiesis}.

\end{itemize}

Attempting to get a handle on anticipatory systems {\em computationally} can 
mean different things to different people.

Suppose, for example, that the mathematical description of the phase web 
described in \S 2 were programmed directly, with all the do-loops, data 
structures, and algorithms this traditionally implies. While the result might 
be 
a good {\em simulation} of an anticipatory a system, I personally would be 
dissatisfied because I seek a system description which {\em by the very nature 
of the computation itself} would produce {\em actual} behavior. That is, while 
the output of such a traditional program is all well and good, the detour 
through an {\em a priori} mathematical description obscures both the mechanism 
and the process by which this output is produced.

Another way to say this is that for me, computation is just as fundamental as 
mathematics, but the two have different strengths. The strength of a 
computational description is that it must exhibit actual {\em mechanisms} and 
the processes engendered thereby. I seek a computational formulation that can 
be 
seen to {\em inevitably} produce systems with the properties listed above, 
without any external or {\em a priori} guiding hand, indeed, with no need to 
appeal to mechanisms beyond what it itself embodies.

This is a tall order! However, I believe I have succeeded to a reasonable 
extent, not least because the resulting purely {\em computational} system- 
descriptive apparatus has (ironically, in view of the preceding comments) a 
very 
clean mathematical formulation (presented in \S 2). Those familiar with the 
various attempts to describe computation mathematically know that the two are 
fractious bedmates, so I view this denoument as a sign that there is something 
very right about it.

In contrast to many, the approach presented here emphasizes {\em structure} so 
strongly that the algorithmic component that for most people is the sine qua 
non 
of computation is nearly non-existent. This emphasis is ultimately the reason 
why the approach offered here - called {\em the phase web paradigm} - differs 
from all others I am familiar with, and correspondingly, why its mathematics 
comes out so differently (algebraic topology, namely, rather than logic).

But how can one even {\em have} computation without an `algorithm'?! The answer 
is that the classical concept of an algorithm is a specification of a {\em 
process} that is to take place when the algorithm is unrolled into time. The 
phase web paradigm is however focused entirely on the process aspect, and 
thereby essentially obviates the need for the {\em a priori} existence of a 
defining algorithm. One might compare this to the theory of evolution based on 
natural selection: this is a process-level theory, for which the existence of 
some {\em a priori} algorithm is problematic.

Of course, one still writes programs, but in pure process terms. However, since 
an anticipatory system in general grows/learns, this programming is ultimately 
sculptural rather than specificational in character.

The next section introduces the basic computational model, which is described 
at 
greater length in [www]. The mathematical translation of this computational 
model follows, and the paper closes by relating all this back to anticipatory 
systems and auto-poiesis.

\newpage
\section{The Computational Model}

The goal of this section is to sketch the essentials of the phase web's 
computational model.

The principal problem computer science has faced over the last two decades is 
the digestion of the phenomenon called ``parallelism'', and virtually all 
contemporary research is colored by issues arising from it. This means that we 
are already in a decidedly {\em process-}oriented context. The computational 
concepts I use - synchronization, co-occurrence, exclusion - are 
well-established and used by researchers in the field. I prefer the term 
``concurrency'' to ``parallelism'' because the latter is tainted by 
associations to interleaving the events constituting several parallel processes 
to achieve a formally sequential process (equivalent to disassembling a living 
cell to form a long end-to-end chain of molecules, and then not even realizing 
that it's dead).

I have been particularly concerned with what are called {\em distributed} 
systems, that is, systems which - like an ant hill - exhibit globally coherent 
behavior via solely local decision-making on the part of its constituents. I 
have been looking for some small set of seed concepts out of which {\em any} 
kind of ``ant hill'' may be built. My goal all along has been to apply the 
understanding gained from this search to construct an entity that can learn 
from its experiences and behave in an increasingly sophisticated way on the 
basis thereof.

As a starting seed, it appears from very general considerations that a 
necessary condition for the ability to profit from experience is 
the ability to draw {\em distinctions}. In a sequential context, this demand is 
met by the {\em if-then-else} construction or equivalent. In the concurrent 
context of the present work, the fundamental distinction I have cooked 
everything down to is that between {\em occur together} versus {\em exclude 
each other}. That is, can two situations co-occur in experience versus they 
cannot self-consistently do so. (The following sub-section therefore treats the 
computational mechanism - synchronization - that addresses such relationships.) 
The overall approach is to express knowledge of self and surround as patterns 
of 
exactly these two {\em complementary} synchronization forms, and to express 
behavior via their manipulation.

The second seed concept is that of symmetry, by which I mean several things:
\begin{itemize}
\item A general symmetry I like is ``outside is as inside'', that is, the {\em 
boundary} separating what is outside from what is inside an entity can be 
drawn arbitrarily, at least in principle. In practice this means that the 
representation of internal relationships should have the same form as the 
representation of external relationships. 

\item A specialization of symmetry is the physicists' use of group-theoretical 
symmetries, which cogently summarize such varied relationships as conservation 
laws, Lorentz (ie. relativistic) invariance, and particle properties. It has 
turned out, though after the fact, as it were, that the phase web's group 
symmetries are very much akin to those of quantum mechanics.

\item A third aspect of symmetry is the requirement that the form of a part of 
a whole is the same as the form of the whole, that is, this is a hierarchical 
requirement. When combined with the ability to harvest observations (cf. {\em 
occur together}), which is a requirement for learning from experience, this 
symmetry leads to the ability {\em internally} to explicitly represent internal 
states and relationships, which in turn supplies the desired self-reflective 
component.
\end{itemize}

The third seed concept is that of goal-directed behavior, by which is meant 
that an entity can explicitly represent to itself the {\em goal} or 
intention of its activity. It is hard to see how this can be avoided; 
the teleological element it introduces is however elastic. Goals can be either 
introduced from the outside or generated internally.

Besides the above concepts, the phase web paradigm is also the product of a two 
broad constraints: `mechanism' and what I call `bio-engineering plausibility'. 
By mechanism is meant that an {\em {\em a priori}} and purely mathematical 
explanation is eschewed in favor of a process-oriented one: the former have 
been 
tried (eg. propositional calculus, Newtonian physics) without particular 
success. The phase web and Topsy are, in contrast, pure process, and this is 
what led to the mathematics we present later, and not the other way around.

By bio-engineering plausibility is meant that the mechanism proposed for a 
computationally-based entity is profitably constrained by requiring that 
this mechanism can conceivably be embodied in biological systems as well. After 
all, the best examples we have of anticipatory systems are biological. The 
information flowing across the boundary from outside the organism to inside 
should, for example, be concrete, should be `grounded': molecular polarity, 
touch, sound waves, retinal pixels, etc. It should perhaps also be noted that 
although a biological system constantly creates and destroys its constituents, 
this is not modelled in the computational model for reasons of efficiency (but 
could otherwise be).

\subsection{Synchronization}

As late as the 1960's main-frame and mini-computers, and again with personal 
computers from the early 1980's until recently, one had {\em one} computer on 
which ran {\em one} program. The coordination between this computer {\em cum} 
program complex and the outside world (ie. ``input/output'') was deeply buried 
in technicalities and generally considered vastly uninteresting. However, when 
one began, with the advent of timesharing, to harbor {\em multiple} programs on 
the same machine, the issue - and profundity - of coordinating the interaction 
of otherwise independent processes gradually became visible.

With multiple interacting processes, a number of new phenomena (at least to 
software people) appeared, eg. concurrency, non-determinism, deadlock, 
communication; and as well, pair of critical new concepts - {\em sharable 
resources} and the necessary {\em mutual exclusion} of processes using same. 
Issues concerned with process interaction and communication came into the 
foreground. All of these things appear in the concurrent world, and none of 
them in the {\em sequential} world of single non-interacting processes.

In order to deal with these things, it was found necessary to introduce a new 
primitive operation into computing, that of {\em synchronization}.\footnote{Not 
to be confused with the synchronization-via-photon-exchange exercises performed 
in relativistic analysis, although the two are of course related.} Viewing an 
`event' as the execution of (say) a single computer instruction, the role of 
computational synchronization is to allow the programmer to specify 
before-after relationships between events belonging to otherwise separate 
processes.

This allows processes that otherwise are unknowing of each other's existence to 
cooperate. Arbitrarily complex inter-process synchronization relationships can 
be built up from primitive before-after relationships. Such synchronization is 
the foundation on which is built all  modern software: your personal computer's 
operating system, local networks, air traffic control, on-line databases, the 
Internet and WWW, $\dots$ everything.

Synchronization possesses a singularly interesting property: it doesn't really 
compute anything! It has the same relationship to the programs that invoke it 
as the pieces in a board game have to the game itself. That is, synchronization 
relationships {\em obtain} while simultaneously being conceptually invisible to 
the processes (ie. game actions) that depend on them. Thus, from the point of 
view of a program, synchronization is not a {\em value}-returning function at 
all, even though textually it often looks like one.
This may be clarified by the following.

{\em \bf Definition.} An {\em event} is a change of state of a system. A 
process 
is a {\em sequence} of such events.

A sequential process with the states $s_{1} \rightarrow s_{2} \rightarrow \dots 
\rightarrow s_{\ell}$ is typically modelled by the composition of {\em 
functions}: $s_{\ell} = f_{\ell}(f_{\ell-1}(\dots f_{2}(f_{1}(s_{1}))\dots)$. 
In 
a typical computational process, the $f_i$ would be arithmetic operations. 
While 
this functional form suffices when there is only one process present (ie. 
traditional programming), analyzing systems with {\em multiple} processes 
encourages the dissolution of this very tight functional binding of states to 
allow us to see the intermediates states as {\em pre-condition, event, 
post-condition}. In this way, the fact that a given pre- or post-condition can 
be caused in more than one way is more readily visible.

The concept of synchronization then allows us to express a multi-process 
computation explicitly in terms of `when' a given pre- or post-condition (ie. 
state) obtains, namely whether before or after (or concurrent with) some other 
state. At this point, the functions $f_{i}$ begin to fade into the background, 
since only their result is visible to other processes. The phase web paradigm 
takes this to its logical extreme: its processes contain {\em no} arithmetic 
functions at all, but rather {\em only} sequences of synchronization 
operations.

The synchronization relationships between processes often possess an invariant, 
which I have argued elsewhere (Manthey,1992) corresponds to a conservation law. 
Conservation laws are group symmetries, not functions. This can be seen as the 
core of the phase web approach, in that the structure, organization, and 
operation of a system is expressed in terms of such invariants. We return to 
this several times in the course of this paper.

By virtue of its before-after focus, synchronization also introduces an 
explicit 
notion of {\em time}, which notion is automatically {\em relative} to events in 
other processes. It is however important to understand that this `time' is 
something much more primitive than that of ordinary usage. [So any decent 
computational theory of physics must build such things as ordinary time (and 
space) up from the relationships obtaining between otherwise isolated primitive 
synchronizations. Conventional theories face their own version of this. I would 
say that I establish plausibility that this is possible in the
phase web.]

Let us now look at the mechanism by which synchronization is 
achieved.\footnote{The story that follows is, at bottom, one of several 
possible 
standard computer science stories, colored by the demands of context.} The two 
operations \Ss{wait} and \Ss{signal} operate on an entity called a `binary 
synchronizer' or `binary semaphore', denoted \Ss{S}. \Ss{S} contains a single 
bit of local state (denoted \Ss{s}) which can take on two mutually exclusive 
values, denoted \Ss{1} and \Ss{\~1}. Define now \Ss{wait} and \Ss{signal} on 
\Ss{S} as follows:

\begin{center}
\begin{sf}
\begin{small}
\begin{tabular}{rlc|cl}
         & S.s=1                  & &   & S.s=\~1 \\ \hline
&&&&\\
wait:    & S.s $\leftarrow$ \~1;  & &   & continue\\
         & return                 & &   & waiting \\
&&&&\\
signal:  & return                 & &   & S.s $\leftarrow $ 1;\\
         &                        & &   & return
\end{tabular}
\end{small}
\end{sf}
\end{center}

The effect of these definitions is to ensure that a given sequential 
computation (ie. process) will stall (namely when \Ss{s=\~1}) until some other 
computation 
\Ss{signal}s it (which sets \Ss{s} to \Ss{1}). Furthermore, a successful 
\Ss{wait} sets \Ss{s} to \Ss{\~1}, thus ensuring that no other computation can 
follow `on its heels'. Notice that
\begin{itemize}
\item no `value' is returned by either operation. Rather, each computation 
simply proceeds on its way after executing \Ss{wait} or \Ss{signal} as if 
nothing had happened;
\item no information is exchanged between \Ss{wait}ing and \Ss{signal}ling 
computations;
\item the effect of the synchronization cannot be `observed' locally (cf. 
preceding item) but will be globally visible as a correlation between events in 
the system as a whole (Manthey,1992);
\item the overall effect is to {\em order} events - namely the respective 
\Ss{wait} and \Ss{signal} events - belonging to two {\em different} 
processes, such that (presuming \Ss{S.s=\~1} initially) the \Ss{wait} in the 
one process will always be after the \Ss{signal} in the other. No more and 
no less.
\end{itemize}

\begin{figure}[htbp]
\begin{center}
\leavevmode
\epsfbox{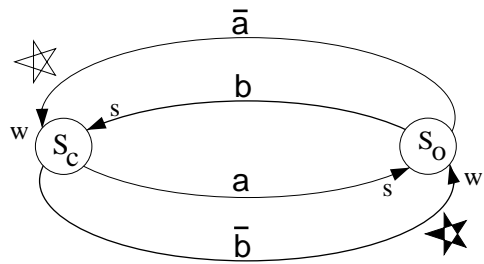}
\caption{\em Synchronization can ensure that certain states (here \Ss{a,b}) 
exclude other. Note that the synchronization stick, initially `in' the 
rightmost 
synchronizer, is conserved.}
\label{Synch}
\end{center}
\end{figure}

These definitions are depicted in Figure \ref{Synch}, in which \Ss{S}$_{o}$ 
({\bf o}pen) corresponds to \Ss{1} and \Ss{S}$_{c}$ ({\bf c}losed) corresponds 
to \Ss{\~1}.
The two processes are denoted by the thick and thin lines, and the two stars 
indicate their starting positions (=states). Following the lines and obeying 
the rules for \Ss{wait} and \Ss{signal}, it it easily seen that state 
\Ss{\{a,\~b\}} excludes state \Ss{\{\~a,b\}}. This state-oriented view is the 
one we take in this paper.

In the figure example, one can conceptualize the alternating mutual exclusion 
between the two computations in terms of a single `synchronization token' - I 
call it a `stick' - that is passed between them like a hot potato. Such a stick 
represents the fact that a particular state obtains. At all times 
there is exactly {\em one} stick present in Figure \ref{Synch}, either in one 
of 
the semaphores or implicitly owned by one of the computations by virtue of the 
state it is currently in. Such a conserved stick, which necessarily must move 
on 
a cyclic (ie. closed) path, reflects the existence of a so-called resource 
invariant.\footnote{ (Manthey,1992) argues the interpretation of this concept 
as 
the computational analog of quantum number conservation laws, and uses it to 
explain how the EPR `paradox' is not a paradox at all.} [Incidentally, the term 
`mutual exclusion' is often abbreviated to `mutex'.]

The preceding discussion has concentrated on the {\em mutual exclusionary} 
effects that can be expressed by synchronization. To express the fact that two 
states can, in contrast, {\em co-occur}, we need only require that the initial 
state of the leftmost synchronizer in Figure \ref{Synch} be {\bf o}pen instead 
of {\bf c}losed. This will allow the co-occurrence of states \Ss{\{a\}} and 
\Ss{\{b\}}, that is, the state \Ss{\{a,b\}} can now occur. [{\em Reader 
exercise}: show that this possibility is unstable or fleeting, and that the 
system can decay into the earlier mutex form. This instability is the lot of 
the 
typical co-occurrence.]

We have thus seen that a synchronizer, which is an archetypic computational 
synchronization mechanism, can be arrayed to express both of the distinctions 
we are after - co-occurrence and exclusion. This particular pair of 
distinctions has the following properties:

\begin{itemize}
\item The elements of a co-occurrence are {\em indistinguishable} in time, in 
that by definition they occur neither before nor after each other. Thus, within 
a co-occurrence there is literally no ``time'' at all: a co-occurrence is a 
``now''.
\item Following Leibniz, co-occurring indistinguishables (namely, 
synchronization sticks) contain the germ of the concept of space. More 
generally, co-occurrence can be extended to encompass such static `structural' 
aspects as form, situation, pattern, and the like.
\item Two successive events of a given process by definition exclude each 
other.\footnote{A consequence of the computational assumption of discreteness. 
Once can rightly say that synchronization is the handmaiden of discreteness.} 
Combining this with viewing ``time'' as a 1-1 mapping of the events 
constituting a given computation to a local time axis, we see that mutual 
exclusion contains the germ of sequential time. In general, every process 
constitutes a local {\em relative} time frame, which frame obtains meaning only 
via synchronization - that is, establishing before-after relationships - with 
other processes' frames.
\item Just as co-occurrence contains the germ of the concept of space, 
exclusion's time-like aspect can be extended to express such dynamic concepts 
as action, transformation, intention, and the like.
\item As a pair, co-occurrence and exclusion over the same states exclude each 
other, thus conceptually closing on each other and leading one to believe that 
they form a complete and minimal set of distinctions.
\end{itemize}

With (Rosen, 1991) in mind, we next investigate a little more closely the 
relationship between synchronization and Turing's model of computation.

\subsection{Escaping from Turing's Box}\label{Co-occ}

An implicit claim of the Turing model is that a single sequence of 
computational 
events can capture all essential aspects of computation, that is, that {\em 
computation consists only of state \underline{transformations}}. To refute 
this claim, consider the following gedanken experiment:

\underline{\em \bf Co-occurrence}

{\em {\bf The coin demonstration} - Act I. A man stands in front of you with 
both hands behind his back, whilst you have one hand extended in front of you, 
palm up. You see the man move one hand from behind his back and place a coin on 
your palm. He then removes the coin with his hand and moves it back behind his 
back. After a brief pause, he again moves his hand from behind his back, places 
what appears to be an identical coin in your palm, and removes it again in the 
same way. He then asks you, ``How many coins do I have?''.}

It is important at the outset to understand that the coins are {\em formally} 
identical: indistinguishable in every respect. If you are not happy with this, 
replace them with electrons or geometric points. Also, I am not trying 
nefariously to slide anything past you, dear reader, in my prose formulation. 
What is at issue is the fact of indistinguishability, and I am simply trying to 
pose a very simple situation where it is indistinguishability, and nothing 
else, 
that is in focus.

The indistinguishability of the coins now agreed, the most inclusive answer to 
the question is ``One or more than one'', an answer that exhausts the universe 
of possibilities given what you have seen, namely {\em at least} one coin. 
There being exactly two possibilities, the outcome can be encoded in one bit of 
information. Put slightly differently, when you learn the answer to the 
question, you will per force have received one bit of information.

{\em {\bf The coin demonstration} - Act II. The man now extends his hand and 
you see that there are two coins in it. [The coins are of course identical.]}

You now know that there are two coins, that is, {\em you have received one bit 
of information.} We have now arrived at the final act in our little drama.

{\em {\bf The coin demonstration} - Act III. The man now asks, ``Where did that 
bit of information come from??''}

Indeed, where {\em did} it come from?! Since the coins are indistinguishable, 
seeing them one at a time will never yield an answer to the question. Rather, 
{\em the bit originates in the simultaneous presence of the two coins}. We have 
called such a confluence a {\em co-occurrence}, and shown how it is computed in 
the preceding section. In that a co-occurrence, by demonstration a bona fide 
computational entity, is `situational' rather than `transformational', the 
assumption that computation is purely transformational is shown to be 
false.

To very briefly dispose of the most common counter-arguments:\\
Q: Whatever you do, it can be simulated on a TM.\\
A: You can't `simulate' co-occurrence sequentially, cf. the coin demo.\\
Q: But you can only check for co-occurrence sequentially - there's always a 
$\Delta t$.\\
A: This is a technological artifact: think instead of constructive/destructive 
interference - a phase difference between two wave states can be expressed 
in one bit.\\
Q: One can simply define a TM that operates on the two states as a whole, so 
the ``problem'' disappears.\\
A: This amounts to an abstraction, which hierarchical shift changes the 
universe of discourse but doesn't resolve the limitation, since one can ask 
this new TM to `see' a co-occurrence at the new level. In general, this type of 
objection dodges the central issue - what is the {\em mechanism} by which 
indistiguishables can be observed.\\
Q: Co-occurrence is primitive in Petri nets, but these are equivalent to finite 
state automata.\\
A: The phase web in effect postulates {\em growing} Petri nets, both in nodes 
and connections. All bets are then off.

[At this juncture, I hasten to mention that we are dealing here with {\em 
local} simultaneity, so there is no collision with relativity theory. Indeed, 
Feynman (Feynman,1965 p.63) argues from the basic principle of relativity of 
motion, and thence Einstein locality, that if {\em anything} is conserved, it 
must be conserved {\em locally}; see also (Phipps),(Pope\& Osborne).

I ought also to mention that I am well 
aware that Penrose (1989) has argued that computational systems, not least 
parallel ditto, {\em in principle} cannot model quantum mechanics. However, I 
believe that his argument, together with most research involving (namely) 
parallelism in my own discipline, is subtly infected with the sequential 
mind-set, going back to Turing's analysis, and truly, earlier. An analogy with 
the difference between Newtonian and 20$^{th}$ century physics is, to my mind, 
entirely defensible. The coin demonstration is my reply to such arguments, 
which I do not then expect to hold.

Notice by the way how the matrix-based formulations of QM neatly get around 
the inherent sequentiality of $y=f(x)$-style (ie. algorithmic) thinking, namely 
by the literal co-occurrence of values in the vectors' and matrices' very 
layouts; and thereafter by how these values are composed {\em simultaneously} 
(conceptually speaking) by matrix operations. Relating this now back to the 
phase web paradigm, if we assign an (arbitrary) ordering on sensor names, then 
co-occurrences become vectors, etc. Instead of the matrix route, I've taken the 
conceptually compatible one of Clifford algebras, which are much more compact, 
elegant, and general, cf. (Hestenes).

Returning to our discussion of Turing's model, we see from the coin 
demonstration that there is information, {\em computational information}, 
available in the universe {\em which {\bf in principle} cannot be obtained 
sequentially.} Thus we have in the coin demonstration a compelling argument 
that, at the very least, the Turing model of computation fails to capture all 
relevant aspects of computation: it is semantically incomplete, and the thing 
it ultimately lacks is {\em space-time} - space: co-occurrence, time: mutual 
exclusion. Synchronization operators represent precisely the way computations 
can express space-time relationships and give them semantic content. 

This can be taken further. Suppose we replace the coins by synchronization 
sticks, which are surely indistinguishable. We can then say that the 
information received from observing a co-occurrence is indicative of the fact 
that two states (represented by their sticks) do not mutually exclude each 
other.

\underline{\em \bf Co-Exclusion}

{\em {\bf The block demonstration}. Imagine two `places', $p$ and $q$, each of 
which can contain a single `block'. Each of the places is equipped with a 
sensor, $s_{p}$ respectively $s_{q}$, which can indicate the presence or 
absence of a block.}

The sensors are the {\em only} source of information about the state of their 
respective places and are assumed {\em a priori} to be independent of each 
other, 
though they may well be correlated. The two states of a given sensor $s$ are 
mutually exclusive, so a place is always either `full', denoted (arbitrarily) 
by $s$, or `empty', denoted by $\tilde{s}$; clearly, $\tilde{\tilde{s}}=s$.

{\em Suppose there is a block on $p$ and none on $q$. This will allow us to 
observe the co-occurrence $\{s_{p}, \tilde{s}_{q}\}$. From this we learn that 
having a block on $p$ does not exclude not having a block on $q$. Suppose at 
some other instant (either before or after the preceding) we observe the 
opposite, namely $\{\tilde{s}_{p}, s_{q}\}$. We now learn that not having a 
block on $p$ does not exclude having a block on $q$. What can we conclude?}

First, it is important to realize that although the story is built around the 
co-occurrences $\{s_{p}, \tilde{s}_{q}\}$ and $\{\tilde{s}_{p}, s_{q}\}$, 
everything we say below applies equally to the `dual' pair of co-occurrences 
$\{s_{p}, s_{q}\}$ and $\{\tilde{s}_{p}, \tilde{s}_{q}\}$. After all, the 
designation of one of a sensor's two values as `$\sim$' is entirely arbitrary. 
It is also important to realize that the places and blocks are story props: all 
we really have is two two-valued sensors reflecting otherwise unknown goings on 
in the surrounding environment. These sensors constitute the {\em boundary} 
between an entity and this environment.

Returning to the question posed, we know that $s_{p}$ excludes $\tilde{s}_{p}$ 
and similarly $s_{q}$ excludes $\tilde{s}_{q}$. Furthermore, we have observed 
the co-occurrence of $s_{p}$ and $\tilde{s}_{q}$ and vice versa. Since the 
respective parts of one co-occurrence exclude their counterparts in the other 
co-occurrence (cf. first sentence), we can conclude that the co-occurrences 
{\em as wholes} exclude each other.

Take this now a step further. The transition $s_{p} \rightarrow \tilde{s}_{p}$ 
is indicative of some {\em action} in the environment, as is the reverse, 
$\tilde{s}_{p} \rightarrow s_{p}$. The same applies to $s_{q}$. Perceive the 
transitions $s_{p} \leftrightarrow \tilde{s}_{p}$ and $s_{q} \leftrightarrow 
\tilde{s}_{q}$ as two sequential computations, each of whose states consists of 
a single value-alternating bit of information. By the independence of sensors, 
these two computations are completely independent of each other. At the same 
time, the logic of the preceding paragraph allows us to infer the existence of 
a third computation, a {\em compound} action, with the state transition 
$\{s_{p},\tilde{s}_{q}\} \leftrightarrow \{\tilde{s}_{p}, s_{q}\}$, denoted 
$s_{p}\tilde{s}_{q}$ or equivalently $\tilde{s}_{p}s_{q}$. In effect, by 
combining in this way two single-bit computations to yield one two-bit 
computation, we have lifted our conception of the actions performable by the 
environment to a new, higher, level of abstraction. This inference we call {\em 
co-exclusion}, and can be applied to co-occurrence pairs of any arity $>1$ 
where at least two corresponding components have changed.\footnote{Greater 
arity is one way to exceed the binary limitation of $\pm 1$ to obtain more 
nuance, though this will not be described further here. Also, the term 
`inference' is to be taken in its generic, not its formal logical, sense: 
co-exclusion is more nearly inductive in its thrust.}

Notice by the way that the same reasoning applies to $\{s_{p},s_{q}\} 
\leftrightarrow \{\tilde{s}_{p},\tilde{s}_{q}\}$, denoted $s_{p}s_{q}$ or 
$\tilde{s}_{p}\tilde{s}_{q}$. The two actions $s_{p}s_{q}$ and 
$\tilde{s}_{p}s_{q}$ are, not surprisingly, {\em dual} to each other, so 
co-exclusion on two sensors can generate two distinct actions. [As will be seen 
later, co-excluding the orientations of the duals produces a ``complete'' 
simplex at the next level up.] Like co-occurrence, an action defined by 
co-exclusion also possesses an emergent property, in this case generally 
comparable to spin $\frac{1}{2}$. This will be made clearer in the mathematical 
discussion below.

It sometimes troubles people that the elements of the co-occurrence (say) 
$\{s_{p}, \tilde{s}_{q}\}$ don't seem at all indistinguishable - on the 
contrary, $s_{p}$ is clearly distinct from $\tilde{s}_{q}$! The confusion is 
understandable, and derives from confounding the {\em value} of a sensor with 
the synchronization {\em stick} that represents the fact that the value (= 
process state) obtains for the moment. The difference is clearer in the 
implementation, where the sticks for the respective states of the sensor 
processes $s_p$ and $s_q$ are represented by the tuples \Ss{[p,1]} and 
\Ss{[q,\~1]}, which tuples can be thought of as making precise exactly {\em 
which} state's stick is being referred to. The processes accessing such tuples 
in fact know {\em a priori} the exact form of the tuple (ie. state) they are 
interested in, so no information is conveyed by accessing such tuples (which is 
as it should be, since synchronization must not convey information between 
processes). Summa summarum, the sensor values are not what are distinguished, 
but rather the sticks representing the associated sensor-process states, and 
these sticks are indistinguishable {\em in time}.

Finally, relative to the co-exclusion inference itself, it provides a very 
general (and novel [Manthey US]) way for an entity to learn from experience: 
simply observe co-excluding co-occurrences, since these then will represent an 
abstraction of experience. Furthermore, this is also neurologically plausible, 
in that co-occurring synapse firings combine to exceed the nerve's threshold. 
The repetition required by neural systems to `remember' is however 
short-circuited in Topsy: once is enough.

\subsection{How Topsy Works}

The trick now is to turn all these observations about co-occurrences and 
co-exclusion-based actions into something that can run on a computer, ie. 
Topsy. 
First, a few general observations:
\begin{itemize}
\item Even though I have made much of true concurrency, it is entirely okay to 
implement Topsy on an ordinary sequential computer, in that one may simply 
accept a certain $\Delta t$ slop in co-occurrence detection. This of course 
means that information deriving from co-occurrences occurring at a granularity 
less than $\Delta t$ will not be available - fair's fair. 

\item It's useful to think of processes as interacting by communicating with 
each other via some medium. In the case at hand, the medium is the computer's 
memory, but it could be wires, micro-waves, QM's spooky action-at-a-distance, 
or whatever. The determining distinction for present purposes is, rather,  
whether a given communication reaches all (``broadcast'') or just a few 
(``point-to-point'') of the other processes. For the phase web paradigm and 
hence Topsy, it is critical that the propagation regime be {\em broadcast}, so 
any process that might be interested in a given synchronization stick, even 
only 
potentially, will have access to it.

\item A very neat way, due to (Raynal), to capture the distinction between 
truly 
distributed system architectures and their imitators  is that whereas the 
imitators implicitly interpret a sent communication as a `request' for 
information and a received communication as a `reply' containing same (which is 
really the same old sequential $y=f(x)$ paradigm disguised as communication), 
processes inhabiting a truly distributed system interpret a communication sent 
as an `announcement' of local state (ie. a stick), and received communications 
as other processes' ditto. Each process decides locally if/when/how it will 
react to the announcements of other processes. The request-reply regime is 
inherently centralizing, whereas the announce-listen regime is inherently 
distributive. It is a fact that virtually all contemporary distributed systems 
are, in this sense, imitators, quite despite appearances.

\item I introduce the concept of a {\em goal} on-the-fly: a goal is an {\em 
explicit} expression of a state that the computation in which it occurs desires 
to reach. Their use in computing goes back to the 1960's in AI (if not 
earlier), and is also found in eg. the language Prolog. Goals may seem unusual, 
since they are at best implicit in traditional `imperative' languages (and also 
in Prolog), but in fact there is nothing new here. Rather, the important thing 
to note is that, by being explicit, goals allow a program using them to 
`remember' what it is supposed to be doing, and thus to recover from blind 
alleys. Furthermore, in being explicit, they allow the program to reason about 
them, and thus eg. reason about and resolve conflicts.
\end{itemize}

Topsy is formally connected to its environment by binary {\em sensors} and {\em 
effectors}, and these together constitute its {\em boundary}. Sensors are 
simple 
two-state processes, which two states are denoted $\{s,\tilde{s}\}$. Effectors 
are viewed as things that influence one or more sensors, and are therefore 
described as $s \rightarrow \tilde{s}$ and vice versa.

Each sensor state is, in the program, converted to a corresponding 
synchronization token, ie. the state $s$ is converted to the token $(s,+1)$, 
and 
$\tilde{s}$ is converted to the token $(s,-1)$. Similarly, if an effector is in 
a state where it carry out the transformation $s \rightarrow \tilde{s}$, this 
is 
converted to the token $(s,+1,-1)$. A goal for this effector would, similarly, 
be expressed by the token $(!,(s,+1),(s,-1))$. In fact, {\em all} program 
states 
of interest are treated like this. In this way, all relationships between the 
processes constituting Topsy can be expressed via synchronization relationships 
alone: there is, as it were, no ``data''... just processes announcing and 
listening for various synchronizational states.

Since an action is defined by co-excluding sensory processes, it expresses both 
a `static' sensor-based aspect - deriving from its defining  pair of 
co-occurrences - and an `active' transformational aspect, deriving from the 
complementarity of these same co-occurrence pairs.\footnote{It would really be 
better to call actions `things', since traditionally a `thing' is namely 
characterized by both aspects. One can also toy with the speculation that 
`syntax' (ie. form) is based on the static, whereas `semantics' (ie. function) 
is based on the active.} These two aspects suggest how to build up a running 
action, namely divide the code for an action into a half devoted to each 
side of the exclusion.

Thus, once the required pair of co-excluding co-occurrences 
$(s_{p},\tilde{s}_{q})$ vs. $(\tilde{s}_{p},s_{q})$ has occurred, a 
multi-threaded\footnote{A {\em thread} is CS jargon for a process possessing a 
relative minimum of own state.} action embodying the two transitions 
$(s_{p},\tilde{s}_{q}) \rightarrow (\tilde{s}_{p},s_{q})$ and 
$(\tilde{s}_{p},s_{q}) \rightarrow (s_{p},\tilde{s}_{q})$, is instantiated 
as a new entity; in a running Topsy system, there will be from hundreds to 
millions of these. One half of an action keys on the co-occurrence  $\{s_{p}, 
\tilde{s}_{q}\}$ and the other on $\{\tilde{s}_{p},s_{q}\}$. Since these 
co-occurrences exclude each other, only one of these halves will be activated 
at a time. When one of these pre-conditions occurs, and at least one associated 
goal is present, the action ``wakes up''. For example, when  $\{s_{p}, 
\tilde{s}_{q}\}$ obtains, along with (say) the goal $s_{p}\rightarrow 
\tilde{s}_{p}$, the action fires and issues a goal for $\tilde{s}_{q} 
\rightarrow s_{q}$ as well. Thus a cascade of transformation goals propagates 
and activates other actions.

Actions carried out at the boundary (effectors) affect the environment, 
causing the sensors to reflect this new situation. This new situation bubbles 
up (see below) through the current aggregration of actions, orienting 
them to the new reality, and old goals are accordingly retracted and new ones 
issued. The seeming anarchy is controlled by the invisible hand of the 
dynamically nested synchronization invariants that the actions represent.

\subsection{The Cycle Hierarchy}

We have now at our disposal co-occurrences, co-exclusion and actions, and 
goals,  and proceed to show how these can be combined recursively to yield a 
hierarchical structure. The basic claim here is that the ability to express the 
complexity and nuance of anticipatory behavior is to be found via the growth 
and 
interplay of hierarchical relationships. This growth, of course, occurs 
naturally and automatically via co-exclusion on sensory experiences.

The hierarchy is called the `cycle hierarchy' because (1) the basic unit of its 
construction is co-excluding processes - the `actions' described above - (2) 
whose internal conservation of synchronization sticks yields a basic cyclic 
structure (cf. Figure 1), (3) which cyclic structure is compounded recursively 
to yield a hierarchy of cycles of cycles.

The cycle hierarchy reflects a {\em weakly} reductionistic stance, in that it 
requires that any higher level phenomenon - which may well be emergent - be 
grounded in the structure and behavior of lower levels. This is in contrast to 
the endemic `subroutine call' or `function composition' hierarchy most people 
(especially scientists and engineers) unconsciously invoke in such discussions. 
This latter hierarchy is {\em strongly} reductionistic, in that it allows {\em 
no} place for phenomena that cannot be modelled by the sequential composition 
of lower level activities.\footnote{To adopt the third possibility, that of 
emergent phenomena in no way grounded in lower levels, is of course to abandon 
any consistent notion of cause and effect and therefore rational thought in 
general. To those readers who see red when the word `emergent' is uttered, I 
note that the concept of emergent phenomena has a counterpart in the  global 
properties found in mathematics, eg. curvature.} The basis of the cycle 
hierarchy in co-occurrences offers an interesting alternative to the reductive 
question of ultimate constituents, namely that one's hierarchical descent 
collides with the boundary to the environment. One is thus ultimately referred 
to ``the rest of the universe'', a result reminiscent of Leibniz's monadology.

Finally, although the following exegesis of the phase web's hierarchical 
structure presumes that the hierarchy is well-nested, ie. like one pancake on 
top of another, this is by no means necessary: co-exclusions can span over 
sensors from multiple levels (Figure 4a is a little misleading in this 
respect). 
Indeed, cycles in the hierarchy itself can be used to express self-propagating 
internal processes.

This overall sketch of hierarchical properties now behind us, we show how such 
hierarchies can be constructed in the first place. The basic insight is:

\begin{itemize}
\item[] {\sc Given} that every action possesses an innate polarity based on the 
orientation of its transformations, $\{s_{p}, s_{q}\} \rightarrow 
\{\tilde{s}_{p},\tilde{s}_{q}\}$ vs. $\{\tilde{s}_{p},\tilde{s}_{q}\} 
\rightarrow \{s_{p},s_{q}\}$, which distinction maps to $\pm 1$, co-occurrences 
of such action polarities can themselves be subjected to the co-exclusion 
inference, producing a meta-level of description/abstraction.
\end{itemize}
In other words, any action, whatever its arity, possesses two locally global 
states, corresponding to the two possible transitions it can accomplish. These 
two states exclude each other, which in turn means that this property of an 
action can be reflected in a two-valued sensor, a so-called {\em meta-}sensor. 
[A meta-sensor is in other respects just like a primitive sensor.]

Meta-sensors themselves can be co-excluded to produce meta-actions, which in 
turn - being, again, actions - possess the same polarities. These 
meta-polarities can again be mapped to a meta-meta-sensor, which can again be 
co-excluded to produce meta-meta-actions, etc. The result is a cycle hierarchy.

Notice that the two complementary co-occurrences whose co-exclusion defines an 
action also neatly specify the respective pre- and post-conditions for that 
action - for example, when the environment is in state 
$\{s_{p},\tilde{s}_{q}\}$, the action's pre-condition is precisely 
$\{s_{p},\tilde{s}_{q}\}$ and its post-condition is $\{\tilde{s}_{p},s_q\}$; 
and vice versa.

When an action's pre-condition obtains, and if a goal to invert (at least) one 
of an action's constituent sensors co-occurs herewith, we say that the action 
is {\em relevant}. The action will then fire, ie. volunteer and broadcast goals 
to invert the actions's remaining constituent sensors, and in so doing attempt 
to achieve said goal from the micro-perspective of that action.\footnote{That 
is, a given co-exclusion, say $\{s_{p},\tilde{s}_{q}\} \leftrightarrow 
\{\tilde{s}_{p},s_{q}\}$, reflects a particularized micro-view of reality that 
says, ``given the goal $s_{p} \rightarrow \tilde{s}_{p}$, if $s_{p}$ is to 
change to $\tilde{s}_{p}$, then this means that $\tilde{s}_{q}$ {\em must} 
change to $s_q$'', and so `volunteer' the goal $\tilde{s}_{q} \rightarrow s_q$, 
which goal, like the first, is visible to all other actions.}

Relevance can be similarly volunteered, on the reasoning ``if $s_{p}$ can be 
changed to $\tilde{s}_{p}$ then an action $\tilde{s}_{p}s_{r}$ can volunteer 
that $s_{r} \rightarrow \tilde{s}_r$ is possible, and therefore is relevant as 
well. Thus volunteering is a way to achieve the associative behavior 
characteristic of anticipatory systems.\footnote{In a traditional frame-based 
AI 
systems, this is called `spreading activation', but it should be apparent that, 
although the effect of the two processes is analogous, the mechanisms are quite 
different.}

Volunteered goals will in general cause other relevant actions to fire, until a 
goal referring to an effector causes that effector to propagate the desired 
effect across the boundary to the environment on the other side thereof. This 
will ultimately change some sensor(s), setting off a wave of changes in the 
associated  relevance relations, reflecting the new state of the environment. 
This interplay between the state of the environment and Topsy's goals occurs 
continually, with current goals changing dynamically in reaction to the 
environment's response to the effects of earlier goals.

Besides volunteering, one other implicit and dynamic mechanism is necessary, 
namely a means for propagating  relevance and goals from level to level. This 
is accomplished by reflecting an action's relevance in an associated 
meta-sensor, whence the same thing will take place for meta-meta-sensors, etc. 
We call this process the {\em bubbling up} of sensory impressions.

Similarly, a goal to invert a meta-sensor will be reflected by the associated 
meta-effector's issuing goals to the level below. Since a given meta-sensor 
represents in one bit the state of a co-occurrence, ie. {\em more than one} 
sensor, a meta-effector fans goals out, level by level, on their way down 
toward 
the primitive effectors at the environmental boundary. We call this process the 
{\em trickling down} of goals.

The hierarchy-construction process leads to a number of features and properties 
deserving mention:

\begin{itemize}
\item The meta-sensors and meta-effectors of a given level form the {\em 
boundary} between that level and the level below. It follows that the boundary 
constituted by the primitive sensors and primitive effectors is, conceptually, 
entirely arbitrary.

\item Since the environment is formally unbounded in its complexity, it follows 
that the hierarchy must be as well. And it {\em is} formally unbounded, in that 
if we abandon the pancake restriction, the number of entities that can be 
co-excluded increases hyper-exponentially: $3, 7, 127, 2^{127}-1$. This is an 
instance of {\em the combinatorial hierarchy} (Bastin and Kilmister), 
(Parker-Rhodes), (Manthey, 1993).

\item Co-exclusion over meta-sensors is inherently introspective and 
self-reflective, in that meta-sensors themselves explicitly express internal, 
situated states. The capture of an internal relationship by a co-exclusion 
elevates what was previously implicit and `unconscious' to an explicit object.

\item We have seen that sensory impressions $S$  bubble up and goals $G$ 
trickle down. A given {\em meta}-level $n+1$ is built over $S_{n}\times S_{n}$, 
and serves to further {\em classify} sensory impressions. When level $n+1$ is 
based only on level $n$, we say the hierarchy is a {\em flat} or {\em pancake} 
hierarchy.

\item One can also consider hierarchies built by co-excluding over $G\times G$ 
and $S\times G$, ie. `meta'-sensors sensing goal-co-occurrences, and 
`meta'-effectors and `meta'-actions manipulating goals for $G\times G$; and 
analogously for $S\times G$.
\begin{itemize}
\item[-] For conciseness, we write (eg.) $G \times G$ for $G \times G \times 
\dots \times G$.
\item[-] $G\times G$, captures relationships between goals, and, 
via hierarchical expansion, can express the structure of arbitrarily complex 
purposive {\em act}ivities. Since $G\times G$ actions are grounded in goals, 
which themselves are primarily internal, their hierarchy is increasingly less 
grounded in environmental reality, wherefore I have dubbed such actions {\em 
icarian}.
\item[-] $S\times G$, expresses the interplay between up-bubbling 
sensory impressions and down-trickling intentions. Since $S\times G$ in a 
sense `covers' both $S\times S$ and $G\times G$, I consider $S\times G$ to be 
the most profound, and call such actions {\em morphic}. Note that morphic 
actions provide a means for expressing the {\em self-}generation of goals given 
sensory situations (read self-choice), and in the other direction, the 
self-generation of sensory situations given goals (read imagination).

Thus the basic phase web mechanisms of co-occurrence and co-exclusion, 
re-applied, can create three distinct {\em types} of hierarchy. In addition, 
entities belonging to each of these can themselves be similarly combined ad 
infinitum. This should provide sufficient expressive power for even the most 
demanding application.
\end{itemize}

\item Bubbling up in an $S\times S$ hierarchy corresponds roughly to 
integration ($\int$), whereas trickling down in the corresponding dual goal 
hierarchy corresponds to differentiation ($\partial$). The $R\times G$ actions 
connecting them correspond then to the meeting of a goal and a currently 
obtaining state, leading to `action'. This is elaborated in the mathematical 
section

\item One can draw an analogy with Huygen's principle as recently elucidated by 
Jessel (Bowden, in press), which says that any radiating primary source, can, 
when surrounded by an arbitrary boundary, be simulated by a finite number of 
appropriately tuned secondary radiators placed on that boundary. Thus 
hierarchical ascent can be compared to approaching the original primary source. 
That the cycle hierarchy is at the same time formally unbounded leads to a 
meta-physically satisfying outcome.
 below.

\item I believe, though without being able to demonstrate it, that moving 
upward in the morphic hierarchy corresponds to a shift to a more powerful 
system in the context of G\"odel's incompleteness arguments.
\end{itemize}

Finally, the initial discriminatory basis for the hierarchy construction - the 
tensions between {\em excludes} and {\em co-occur} and {\em co-exclude} in {\em 
time} - seems to blur in their interplay the traditional distinction between 
epistemology and ontology. This obtains because, while co-occurrences and 
co-exclusion-based actions together constitute the universe of `ontological 
objects', their discovery (ie. epistomology) invokes the very same properties. 
Only after one has built up considerable structure - corresponding to 
traditional space-time - would one seem to be able to clearly separate the two.

\section{The Mathematical Model}

This section presents, very informally, the most important mathematical aspects 
of the phase web paradigm. In general, the vector orientation of the present 
approach is unique in computing, which has traditionally been 
logic-oriented.

\begin{figure}[htbp]
\begin{center}
\leavevmode
\epsfbox{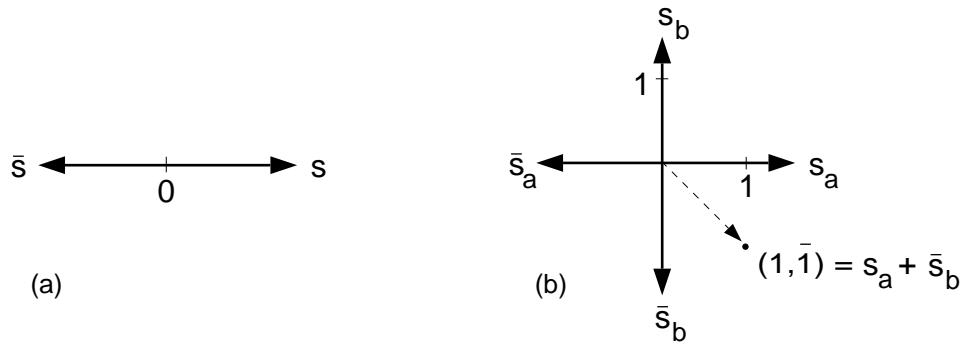}
\caption{Sensors as vectors.}
\label{SensorAsVect}
\end{center}
\end{figure}

The point of departure is to view sensor states as vectors instead of scalars, 
as is conventionally done.\footnote{A {\em scalar} is simply a magnitude, 
whereas a {\em vector} is a magnitude together with a direction (orientation). 
The operations on vectors ($+, \cdot, \wedge$) ensure that one's intuitive 
expectations for how things combine are maintained.} Figure \ref{SensorAsVect}a 
shows a single sensor's states so expressed, and Figure \ref{SensorAsVect}b the 
way two such vectors can indicate a state, eg. of an action.

The sensor state $s=1$ indicates that sensor $s$ is  currently being 
stimulated, ie. a synchronization stick for that state is present, whereas 
$s=\tilde{1}$ indicates that $s$ is currently {\em not} being stimulated, and 
hence no stick for state $s$ is present. Thus the two states of $s$ are 
represented by the respective semaphore values introduced in the definition of 
\Ss{wait} and \Ss{signal} in \S 1.1.

That the sensors {\em qua} vectors are orthogonal derives from the fact that, 
in 
principle, a given sensor says nothing about the state of any other sensor. A 
state of a multi-process system such as that depicted in Figure 
\ref{SensorAsVect}b is then naturally expressed as the sum of the individual 
sensor vectors. For example, the state $(s_{a},\tilde{s}_{b}) = (1,\tilde{1})$ 
is written as the vector sum $s_{a}+\tilde{s}_{b}$, which also introduces the 
visual convention that a vector component written without a tilde is taken to 
be bound to the value $1$, and vice versa. Since such states represent 
co-occurrences, it follows that co-occurrences are vector sums. Note how the 
commutativity of `$+$' reflects the lack of ordering of the components of a 
co-occurrence.

The next step is to find a way to represent actions mathematically. 
(Manthey,1994) presents a detailed analysis of the group properties of both 
co-occurrences and actions, concluding that the appropriate algebraic formalism 
is a (discrete) Clifford algebra, and that the state transformation effected by 
an action is naturally expressed using this algebra's vector product. A prime 
characteristic of this product is that it is anti-commutative, that is, for 
$(s_{1})^2 = (s_{2})^2 = 1$, $s_{1}s_{2} = -s_{2}s_{1}$.\footnote{The Clifford 
product $ab$ can be defined as $ab=a\cdot b + a\wedge b$, ie. the sum of the 
inner ($\cdot$) and outer ($\wedge$) products, where $a\wedge b = -b\wedge a$ 
is the oriented area spanned by $a,b$. The vector cross product $a\times b$ 
familiar to many is a poor man's version of $a\wedge b$ introduced by Gibbs. 
The basis vectors $s_{i}$ of a Clifford algebra may have $(s_{i})^2=\pm 1$, and 
while here we choose $+1$, reasons are appearing for choosing $-1$. As long as 
they all have the same square, it doesn't matter for what is said here.} The 
magnitude of any such product is the area of the parallelogram its two 
components span, and the {\em orientation} of the product is perpendicular to 
the plane of the parallelogram and determined by the ``right hand rule''.

Applying the Clifford product to a state, one finds - using the square-rule and 
the anti-commutativity of the product given above - that 
\begin{equation}
(s_{1}+s_{2})s_{1}s_{2} = s_{1}s_{1}s_{2} + s_{2}s_{1}s_{2} = s_{2} + 
\tilde{s}_{1}s_{2}s_{2} = \tilde{s}_{1} + s_{2}
\label{(1)}
\end{equation} 
that is, that the result is to rotate the original state by $90^{o}$, for which 
reason things like $s_{1}s_{2}$ are called {\em spinors}. Thus {\em state 
change} in the phase web is modelled by rotation (and reflection) of the state 
space, and the effect of an `entire' action can be expressed by the inner 
automorphism $s_{1}s_{2}(s_{1}+s_{2})s_{2}s_{1} = \tilde{s}_{1}+\tilde{s}_{2}$, 
which corresponds to a rotation through $180^o$.\footnote{Some readers might 
recognize this when written in the form $as=s^{\prime}a$.} It is interesting to 
note that $(s_{1}s_{2})^2=-1$, that is, the $s_{1},s_{2}$-plane is the 
so-called {\em complex} plane, and thus that $i = \surd -1$ is intimately 
involved.

One of the felicities of Clifford algebras is that one needn't designate one of 
the axes as `imaginary' and the other as `real'. Rather, the $i$-business is 
implicit  and the  algebra's anti-commutative product  neatly bookkeeps the 
desired orthogonality and inversion relationships. 

The above spinors are just one example of the vector products available in a 
Clifford algebra - any product of the basis vectors $s_i$ is well-defined, and 
just as $s_{1}s_{2}$ defines an area, $s_{1}s_{2}s_{3}$ defines a volume, etc.  
Not least because they are all by nature mutually perpendicular, the terms of a 
Clifford algebra 
\begin{equation}
s_{i} + s_{i}s_{j} +  s_{i}s_{j}s_{k} + \dots + 
s_{i}s_{j}\dots s_{n}
\label{(2)}
\end{equation} 
themselves also define a vector space, which is the space in which we will be 
working. [The term (eg.) $s_{i}s_{j}$ above, for $n=3$, denotes $s_{1}s_{2}+ 
s_{2}s_{3}+s_{1}s_{3}$, that is, all possible non-redundant combinations.]

At this point it is perhaps worth stressing that this vector space is the space 
of the {\em distinctions} expressed by sensors, and as such has no direct 
relationship whatsoever with ordinary 3+1 dimensional space. The latter must - 
at least in principle - be built up from the primitive distinctions afforded by 
the sensors at hand. This too is treated as a discrete space, rather than the 
usual continuous ditto.

A Clifford product like $s_{1}s_{2}$ reflects both the emergent aspect of a 
phase web action (via its perpendicularity to its components) and its ability 
to act as a meta-sensor (since its orientation is $\pm 1$).

One might therefore expect that the co-exclusion of two such meta-sensors, say 
$s_{i}s_{j}$ and $s_{p}s_{q}$, would be modelled by simply multiplying them, to 
get the 4-action $s_{i}s_{j}s_{p}s_{q}$. This turns out however to be 
inadequate, since although by the same logic the co-exclusion of (say) $s_{i}$ 
and $s_{i}s_{j}$ in Topsy expresses explicitly a useful relationship (eg. 
part-whole), the algebra's rules reduce it from $s_{i}s_{i}s_{j}$ to $s_j$, 
which is simply redundant.

Instead, we take as a clue the fact that goal-based {\em change} in Topsy 
occurs via trickling down through the layers of hierarchy, and draw an analogy 
with differentiation. In the present decidedly geometric and discrete context, 
differentiation corresponds to the {\em boundary operator} $\partial$. 
Informally, define $\partial s = 1$ and let
$$\partial(s_{1}s_{2}\dots s_{m}) = 
s_{2}s_{3}\dots s_{m} - s_{1}s_{3}\dots s_{m} + s_{1}s_{2}s_{4}\dots s_{m} - 
\dots (-1)^{m+1}s_{1}s_{2}\dots s_{m-1}$$ 
that is, drop one component at a time, in order, and alternate the sign. Using 
the algebra's rules, one can show that 
$$\partial(s_{1}s_{2}\dots s_{m}) = 
(s_{1}+s_{2}+\dots+s_{m})s_{1}s_{2}\dots s_{m}$$
which is exactly the form of equation (1) for what an action does.

The boundary operator $\partial$ has a straightforward geometric 
interpretation. Consider an ordinary triangle $ABC$ specified in terms of its 
vertices $A,B,C$, whence its edges are $AB, BC, CA$. Then \[\partial(ABC) = BC 
- AC + AB\] Since specifying the triangle's edges in terms of its vertices 
means that edge $AC$ is oriented oppositely to edge $CA$, we can rewrite the 
above as $AB + BC + CA$, which is indeed the boundary of the triangle (versus 
its interior).

To find co-exclusion in this, we exploit the geometrical connection further. 
Take exppression (2) expressing the vector space of distinctions, segregate 
terms with the same number of product-components ({\em arity}), and arrange 
them 
as a decreasing series:
\begin{equation}
 s_{i} \stackrel{\partial}{\longleftarrow} s_{i}s_{j} 
\stackrel{\partial}{\longleftarrow} s_{i}s_{j}s_{k} 
\stackrel{\partial}{\longleftarrow} \dots \stackrel{\partial}{\longleftarrow}  
s_{i}s_{j}\dots s_{n-1} \stackrel{\partial}{\longleftarrow} s_{i}s_{j}\dots 
s_{n}
\label{(3)}
\end{equation}
Here as before, $s_{i}s_{j}$ is to be understood as expressing all the 
possible 2-ary forms (etc.), and hence the co-occurrence of pieces of 
similar structure. Each of the individuals is a {\em simplicial complex}, and 
the whole mess is called a {\em chain complex}, expressing a sequence of 
structures of graded geometrical complexity in which the transition from a 
higher to a lower grade is defined by $\partial$. Furthermore, the entities at 
adjacent levels are related via their group properties - their {\em homology}, 
which I here assume is trivial.

Still on the scent of co-exclusion, it turns out that there is a second 
structure - a {\em cohomology} - that is isomorphic to (``same {\em form} as'') 
the homology, but with the difference that arity (complexity) {\em increases} 
via the $\delta$ (or {\em co-boundary}) operator,\footnote{More precisely, 
$(\sigma_{p},\delta d^{p-1})=(\sigma_{p}\partial,d^{p-1})$, where $\sigma_{p}$ 
is a simplicial complex with arity $p$, and $d^{p}$ the corresponding 
co-complex.} precisely opposite to $\partial$ (cf. equation (3)):
\begin{equation}
s_{i} 
\stackrel{\delta}{\longrightarrow} s_{i}s_{j} 
\stackrel{\delta}{\longrightarrow} s_{i}s_{j}s_{k} 
\stackrel{\delta}{\longrightarrow} \dots \stackrel{\delta}{\longrightarrow}  
s_{i}s_{j}\dots s_{n-1} \stackrel{\delta}{\longrightarrow} s_{i}s_{j}\dots 
s_{n}
\label{(4)}
\end{equation}
Building such increasing complexity is exactly what co-exclusion does. [I note 
that a Clifford algebra satisfies the formal requirements for the existence of 
the associated homology and cohomology.]

\begin{figure}[htbp]
\begin{center}
\leavevmode
\epsfbox{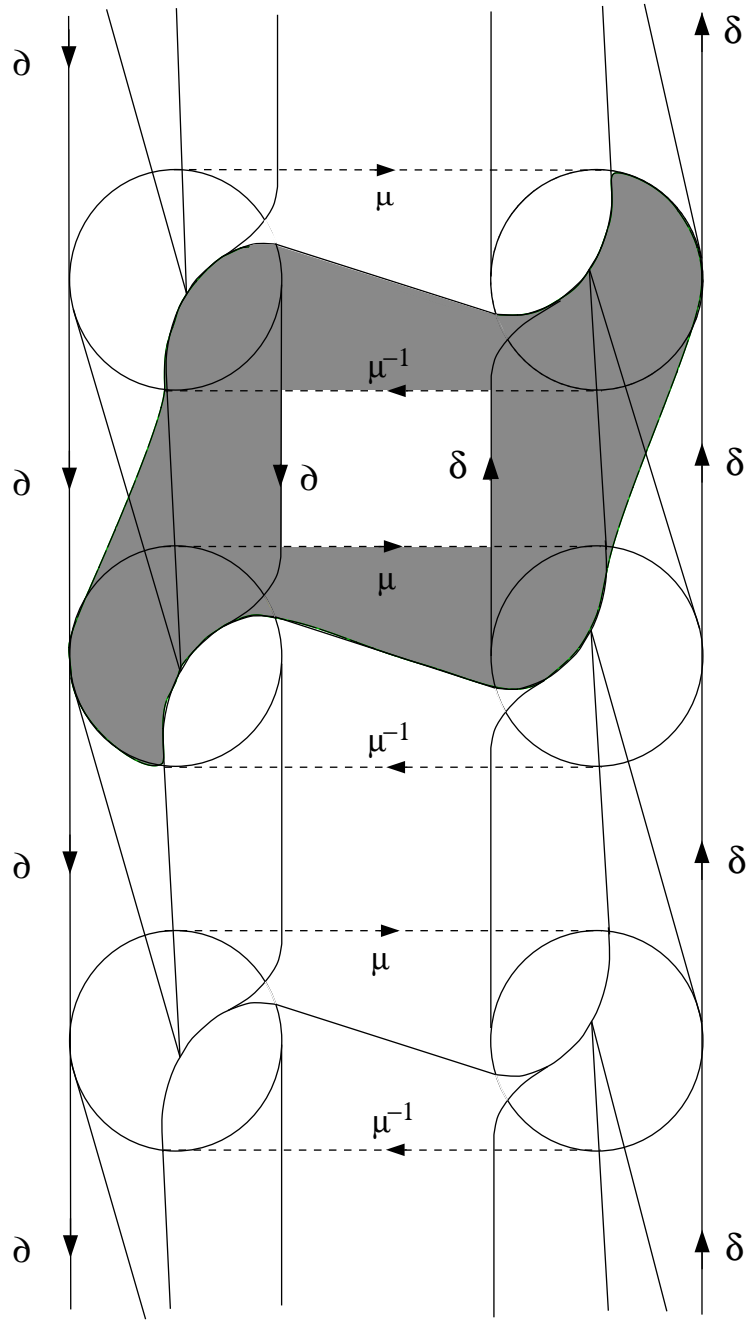}
\caption{Ladder diagram, illustrating homology-cohomology relationships.}
\label{BowdenBase}
\end{center}
\end{figure}

Figure \ref{BowdenBase}, due to (Bowden,1982), illustrates these relationships 
(eqns. 3,4). I call this a {\em ladder diagram}.

The left side of the ladder is the homology sequence generated by $\partial$ 
over the representation of actions as Clifford products. The downward flow of 
decomposition of the structure into simpler pieces (ie. the crossing of 
successive boundaries) corresponds to the trickling down of goals described 
earlier.

The right side of the ladder is similarly the cohomology sequence generated by 
$\delta$ from sensory impressions. The upward flow of composition of structure 
to form more complex structure corresponds to the effect of co-exclusion, up 
through which increasingly complex structure sensory impressions bubble.

The circles represent all the entities (Clifford algebra terms) at the 
particular level of complexity. The larger of the two circle-halves holds those 
entities which will map to zero with the next hierarchical transition 
($\partial$ or $\delta$) - called the {\em kernel} of the group - as indicated 
by the pointed `beak'.

The rungs of the ladder, besides denoting the location and content of cycle 
hierarchy levels, also express the fact that there exist isomorphisms 
($\mu,\mu^{-1}$) between the structures at either end of a given 
rung.\footnote{Strictly speaking, $\mu/\mu^{-1}$ should be indexed by level, 
$\mu_{m}/\mu^{-1}_{m}$.} The
shaded portion, which can be seen to repeat in both directions, expresses the 
so-called commutation relationships that obtain. That is, if one chooses a 
particular group element and follows the transforming arrows around the 
interior box, one not only arrives back where one began, but also back at the 
exact same {\em element} one began with! One says that the isomorphisms {\em 
commute}, and one may also take longer paths, though always obeying the 
box-arrows (otherwise the commutation relation generally won't hold).

The shaded shape points out a unique property of the homology-cohomology 
ladder, one that even most topologists seem unaware of, namely that the 
isomorphisms $\mu, \mu^{-1}$ are {\em twisted}, that is, the kernel of the 
group at one end of a rung is mapped by $\mu$ (respectively, $\mu^{-1}$) into 
the non-kernel elements of the group at the other end. This property was 
discovered by (Roth) in his proof of the correctness of Gabriel Kron's (then 
controversial) methods for analyzing electrical circuits (Bowden), and turns 
out to have profound implications. For example, the entirety of Maxwell's 
equations and their interrelationships can be expressed by a ladder with two 
rungs plus four terminating end-nodes (Bowden), and (Tonti) has - 
independently - shown similar relationships for electromagnetism and 
relativistic gravitational theory. Roth's twisted isomorphism (his term) thus 
reveals the deep structure of the concept of boundary, and shows that the 
complete story requires both homology and cohomology.

I interpret the twisted isomorphism to be expressing a deep complementarity 
between the concepts of action and state, between exclusion and co-occurrence. 
In the running Topsy program, $\mu,\mu^{-1}$ connect goals' trickling down to 
sensory states' bubbling up. Think now about this: suppose you follow only the 
goal/homology side down. As boundary after boundary is crossed, all that 
happens is that a larger goal is split into successively narrower subsidiary 
goals. Imagine that you are following a ladder structure that describes the 
entire Universe. When and where does the actual {\em change} occur?!

The answer is that it never does, as long as you stick to the homology side of 
the ladder. Similarly, if you stick to the cohomology side, states never 
turn into goals: there is eternal stasis. It is the mappings $\mu,\mu^{-1}$ 
that allow dynamic and change, converting something that doesn't exist, even 
conceptually, on the one side to something that constitutes the conceptual 
universe of the other. But this is what morphic actions do explicitly, so 
$\mu,\mu^{-1}$ correspond to actions over $S\times G$.\footnote{Given that I 
associate wave properties with the concept of co-occurrence and particle 
properties with exclusion and action, this means that $\mu,\mu^{-1}$ express 
wave-particle duality, but in a hierarchical structure that itself expresses no 
difference between the microscopic and the macroscopic.}

\begin{figure}[htbp]
\begin{center}
\leavevmode
\epsfbox{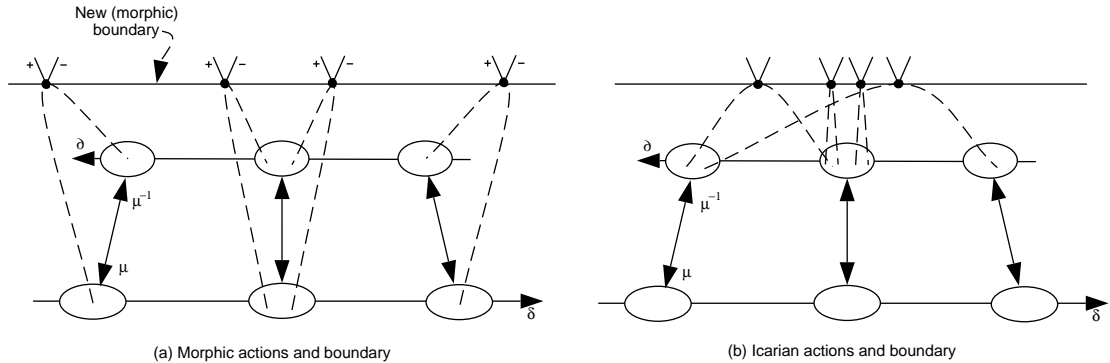}
\caption{Morphic vs. icarian hierarchies.}
\label{MIHier1}
\end{center}
\end{figure}

Summarizing, we have seen how morphic actions correspond to $\mu,\mu^{-1}$, and 
it should be clear that the conversion of meta-sensors to 
meta-actions\footnote{Linguistic and conceptual purity would demand, since 
we're 
on the $\delta$/state side of the ladder, that I write `meta-object' or 
`meta-state'  instead of `meta-action', but this distinction is blurred outside 
of a mathematical context, so I don't.} via co-exclusion ($S_{n}\times S_{n}$) 
corresponds to $\delta$; similarly, icarian actions ($G_{n} \times G_{n}$) 
correspond to $\partial$; non-pancake hierarchies are, of course, more complex 
mathematically. Figure \ref{MIHier1} illustrates two possibilities, both 
non-pancake. 

In conclusion, the attempt to put the phase web and Topsy on a firmer 
mathematical footing turns out to lead to the same mathematical structures as 
underpin both contemporary physical and (I understand) bio-physical theory. 
Things could hardly have turned out better for a novel computational approach 
to 
expressing information, learning, behavior, and in general, the structures and 
mechanisms pertinent to anticipatory systems.

\section{Modelling Auto-Poiesis}

Our goal now is to describe auto-poiesis (Maturana and Varela, 1987) using the 
apparatus of the phase web. We have two ways to express the latter at hand, a 
computational way and mathematical way. The former provides many important 
details that the latter lacks, but since this lack is an advantage in the 
present context, we will couch the description with reference to the 
mathematical version.

Auto-poesis invokes three interdependent activities: (1) sensing the 
environment 
and categorizing the information so gained; (2) acting on this information in a 
manner consistent with the structure of said environment; and finally, (3) 
acting in such a way as to be able to maintain, and even improve, the ability 
to 
carry out these three activities.

It should be stressed that the unpredictability of the environment effectively 
precludes a completely pre-programmed solution, if for no other reason than 
that 
the combinatorial complexity of the impulses coming from the environment 
together with the desired response, if explicitly listed, would simply take up 
too much `room'; that is, such a solution is, from a bio-engineering point of 
view, implausible. This conclusion obtains even when, as is the case with (say) 
insects, the top-level behavior is apparently very rigid, since even walking 
about or flying cannot be similarly rigid. Accordingly, we assume that the 
solution must be non-deterministic and adaptive in character.

We now take up each of the above three activities.

\subsection{Sensing and Categorizing}

The key problem here is the sheer combinatorics of the sensory input: $n$ 
sensors taken $1,2,3\dots n$ at a time yields $2^n$ possible combinations. 
Since 
even a simple cell can be construed to interact with its environment over most 
of its surface, clearly $n$ is very large. Moreover, this analysis does not 
consider that it is not least the {\em order} in which the various sensory 
combinations occur that is important, in which case things grow factorially.

The concept of {\em hierarchy} is therefore a critical bio-engineering {\em 
tool}. This is so because, as already noted by (Simon, 1967), a hierarchical 
organization reduces complexity logarithmically. However, even Simon's analysis 
implicitly invokes a subroutine-call/functional hierarchy, and overlooks the 
fact that the multiplicity of simultaneous impulses from the environment 
requires multiple, likely overlapping, hierarchies, one for each context so 
invoked.

This is where the bubbling-up of sensory impressions characteristic of the 
cycle 
hierarchy (ie. the $\delta$-side of the ladder) comes to the rescue: the 
$\delta$ structure contains implicitly all {\em possible} 
hierarchies\footnote{Modulo, of course, the distinctions the organism is in 
fact 
capable of.}, and the bubbling-up process orients the structure to {\em which} 
hierarchies are relevant in the given context. The particular orientation thus 
achieved is at the same time {\em ipso facto} a {\em categorization} of the the 
sensory stimuli in that context.

Therefore, we model the categorization of an organism's sensings as a simple 
meta-hierarchy, where each level builds essentially directly on the level 
below. 
How many such levels there might be for this is presumably dependent on the 
organism's complexity. The top level of such a hierarchy will contain the 
highest-level categorizations, and presumably most contexts will ``light up'' 
several, though distinct, top-level category nodes.

\subsection{Environmentally Appropriate Behavior}

Once these categorizations of the environment are available, we can address the 
issue of choosing an appropriate response. A given response can in general be 
expected to span several modalities (ie. distinct ``effectors'').

Even though the response is not required to be especially prescient (this is 
the 
responsibility of (3) above), it is nevertheless a fact that even the most 
trivial `intelligent action' requires a fair amount of organization and 
coordination to achieve. For example, bio-engineering economy requires that the 
same components be used to respond to similar situations, and ensuring that 
this 
functional overlap does not get in the way of the correct response in the {\em 
particular} situation leads to the need for the aforementioned coordination. 
The 
example in (Manthey,1996) illustrates this clearly.

Another way to put this is that, given that the categorization process selects 
one or several categories out of many, the generation of appropriate behavior 
can be characterized as the ability to control the transition from the current 
category-set to an intended new ditto. This in turn requires a structure that 
spans over the existing categorization structure. We write `span over' (when 
`span' alone would have been sufficient) to emphasize that we are not speaking 
here of simply adding levels on top of the categorization meta- hierarchy, but 
rather a {\em morphic} hierarchy (cf. eg. Figure \ref{MIHier1}a) that literally 
spans {\em over} the entire categorization structure, more or less from top to 
bottom.

This superstructure, besides exhibiting clearly the greater flexibility and 
generality of the ladder hierarchy compared to functional ditto, is genuinely 
self-reflective. This obtains, as should be clear from the figure, because it 
is 
the mechanism and dynamics of the categorization process {\em itself} that are 
being abstracted over.

We choose a morphic hierarchy primarily because a meta-hierarchy (which might 
otherwise be considered as a candidate here) cannot itself generate its own 
goals (being constructed over $S \times S$), whereas this is a principle 
characteristic of a morphic hierararchy (which is constructed over $S \times 
G$). The morphic superstructure, consisting (say) of two-four levels, has the 
responsibility for issuing goals to the underlying meta-hierarchy. It is these 
goals that control the transition from the current situation to a desired new 
one. [In this connection, it is perhaps appropriate to emphasize that the 
structures we are describing are entirely comfortable with unexpected reactions 
to their effector-born manipulations from the environment: the goal-driven 
regime ensures that the organism will adaptively pursue the achievement of its 
intents in the face of non-deterministic outcomes.]

\subsection{Auto-Poietic Behavior}

The goal of auto-poietic behavior is to ensure the continued existence of the 
organism. To this end, it seems obvious to simply iterate the logic of the 
preceding construction. That is, just as we above used a spanning hierarchy to 
self-reflect the categorization process in order to control the transitions 
between categorizations, we now introduce a second spanning hierarchy to 
self-reflect over the transition-control process.

This second spanning hierarchy must, like the preceding, be able to generate 
goals, although in this case the goals are meant to control the long-term 
behavior of the organism. That is, it is entirely conceivable that in all but 
the most primitive organisms, there are a number of more or less mutually 
exclusive reactions that could be generated in a given situation. The preceding 
morphic hierarchy cannot `intelligently' choose among these possibilities 
because their mutually exclusive properties are not explicitly visible to 
it(self). Thus, another way to describe the utility of the superstructures we 
are building here is that they make explicit various relationships that are 
entirely implicit in the structure over which they brood.

Whether this second self-reflective level should be morphic or icarian is at 
this point a matter of speculation. Icarian structures, because they are built 
solely over $G \times G$, that is the co-occurrence or mutual exclusion of 
goals 
themselves, are naturally suited to controlling longer sequences of actions, 
ie. 
actions that must be carried out in a {\em particular} order in order for the 
whole sequence to be successful. To accomplish this, icarian actions interact 
with their root boundary by retracting and (perhaps later, re-)issuing goals 
`belonging' to the underlying hierarchy, here the morphic one just described. 
In 
that goals are purely internal to the organism, the danger exists that a deep 
icarian hierarchy can become so involved in pursuing its own goals that it 
loses 
sight of the actual environmental situation and feedback. In contrast, a 
morphic 
hierarchy avoids this - it's by definition (more) directly connected to the 
environment via its $S$-component - at the expense of tending toward hard-wired 
reaction and a weaker ability to manage the interplay between goals.

In either case, one can argue that this auto-poietic superstructure should be 
rooted not only in the underlying morphic hierarchy but also in the original 
`primitive' level. This will help  the overall structure to maintain a closer 
connection to environmental reality. One can also consider literally 
feeding-back the results of this stage into the lower level(s), again in the 
errand of greater cohesion of the whole.

Of course, there is still no guarantee that this second self-reflective level 
will succeed in the long term, and the same applies no matter how many such 
levels exist, so at some point, one must appeal to natural selection and its 
accompanying phenomena for the final auto-poietic judgement. 

Howsoever, we claim that for an organism to display auto-poietic behavior, it 
must possess, as a minimum, the above-described three level-structures - one to 
categorize the current state of the environment vis a vis the organism, one to 
connect these categorizations to immediate reactions, ``tactics'', and one to 
oversee  its long-term behavior, ``strategy''.

\section{Summary and Conclusion}

We have presented a situated computational model - the {\em phase web} - that 
we 
believe capable of describing the true structure and behavior of anticipatory 
systems. It is a {\em pure process} model whose fundamental departure from 
traditional algorithmic thinking allows it to meet Rosen's (1991) criticism of 
the latter, and which can express emergent phenomena. Being a {\em 
computational} model, it is able to propose explicit mechanisms and processes 
for acquiring and using, adaptively, information from its environment. The key 
insight is to express the desired activities in terms of {\em patterns of 
synchronization} among the events constituting an organism. These patterns fall 
into two distinct categories - co-occurrence and mutual-exclusion - that 
together are capable of expressing the concepts of event (synchronization 
itself), process (via exclusion), information (cf. the coin demonstration), 
space (via co-occurrence), time (via exclusion), action (via co-exclusion), 
structure (via hierarchy), self-reflection (via co-occurrence and exclusion 
over 
internal events), intent (via goals), etc. As a result, the model need not 
appeal to mechanisms outside of itself, that is, the modelling tools themselves 
exhibit logical closure, and are, apparently, complete.

After a brief description of how a program embodying these concepts (ie. Topsy) 
actually works, we showed how this same model can also be described in terms of 
algebraic topology. The key identifications making this possible are: (1) a 
binary sensor can be viewed as a vector, and the set of sensors connecting an 
system to its environment as an orthonormal basis; (2) the sum of sensor 
vectors 
captures the concept of co-occurrence and their (Clifford) product the concepts 
of exclusion and action; (3) the co-exclusionary property of complementary 
co-occurrences allows the composition of sensory-object abstractions from 
environmental stimulation and corresponds to the co-boundary operator $\delta$; 
(4) this induces a hierarchy of co-boundaries and co-chain complexes, which in 
turn, via Roth's twisted isomorphism, (5) induces a corresponding and {\em 
isomorphic} homology - ie. a hierarchy of boundaries and chain complexes - that 
expresses the {\em de}composition (via the boundary operator $\partial$) of 
goals on meta-sensors into sub-goals/sub-actions, leading ultimately to 
externally-directed effects on the environment; and finally, (6) this {\em 
ladder hierarchy} yields three distinct types - meta, morphic, and icarian - 
corrresponding (very roughly) to classification, situated reaction, and 
goal-interaction.

Finally, the preceding section showed how to apply this model to the 
description 
of auto-poietic behavior. We concluded that an auto-poietic system must as a 
minimum contain three distinct hierarchies, one to classify sensory input in a 
(presumed) non-deterministic regime; one to self-reflectively react to the 
current, now classified, situation in an appropriate and controlled manner; and 
one to self-reflectively choose among the possible, now identified, reactions, 
and in so doing pursue the overall goal of achieving and maintaining 
auto-poiesis.

We claim as well that this cannot be achieved within reasonable bio-engineering 
constraints without invoking the ladder-hierarchical structures we have 
described. We claim further that these hierarchical structures must be 
interconnected essentially as described in order to obtain the 
self-reflectivity 
without which very little of this behavior can be achieved.

Finally, we claim that these are not burdensome constraints at all, but rather 
just those that are needed, both in terms of modelling apparatus and technique, 
to describe that which is so very, very special about anticipatory systems.

\section*{Acknowledgements}

Thanks to IEEE for permission to reprint the coin and block demonstrations. And 
special thanks to Peter Marcer, Arne Skou, and my students.

\vspace{0.5cm}
\section*{References}

Bastin, T. and Kilmister, C.W. {\em Combinatorial Physics}. World Scientific, 
1995. Isbn 981-02-2212-2.

Bastin, T. and Kilmister, C.W. ``The Combinatorial Hierarchy and Discrete 
Physics''. Int'l J. of General Systems, special issue on physical theories from 
information (in press).

Bowden, K. ``Physical Computation and Parallelism (Constructive Post-Modern 
Physics)''. Int'l J. of General Systems, special issue on physical theories 
from information (in press).

Bowden, K. {\em Homological Structure of Optimal Systems}. PhD Thesis, 
Department of Control Engineering, Sheffield University UK. 1982.

Feynman, R. {\em The Character of Physical Law}. British Broadcasting Corp. 
1965.

Gelernter, D. ``Generative Communication in Linda''. ACM Transactions on 
Programming Languages, 1985.

Hestenes, D. and Sobczyk, G. {\em From Clifford Algebra to Geometric Calculus.} 
Reidel, 1989.

Hestenes, D. {\em New Foundations for Classical Mechanics}. Reidel, 1986. The 
first 40 pages contain a very nice, historical introduction to the vector 
concept and Clifford algebras.

Manthey, M. ``Synchronization: The Mechanism of Conservation Laws''. Physics 
Essays (5)2, 1992. 

Manthey, M. ``The Combinatorial Hierarchy Recapitulated''. Proc. ANPA 15 
(Cambridge, England), 1993.

Manthey, M. ``Toward an Information Mechanics''. Proceedings of the 3rd IEEE 
Workshop on Physics and Computation; D. Matzke, Ed. Dallas, November 1994. Isbn 
0-8186-6715X.

Manthey, M. et al. Topsy - A New Kind of Planner. Working paper, 1996. See 
[www].

Manthey, M. US Patents 4,829,450, 5,367,449, and others pending. My intent 
is to license freely (on request) to individuals  and research or educational 
institutions for non-profit use. See [www] for licensing information.

Maturana, H.B. and Varela, F.J. {\em The Tree of Knowledge - the biological 
roots of human understanding}. New Science Library; Shambhala, 1987. Isbn 
0-87773-403-8.

Phipps, T. ``Proper Time Synchronization''. Foundations of Physics, vol 21(9).  
September, 1991.

Penrose, R. {\em The Emperor's New Mind}. Oxford University Press, 1989. Isbn 
0-19-851973-7.

Pope, N.V. and Osborne, A.D. ``Instantaneous Relativistic 
Action-at-a-Distance''. Physics Essays, 5(3) 1992, pp. 409-420.

Parker-Rhodes, F. {\em The Theory of Indistinguishables - A Search for 
Explanatory Principles Below the Level of Physics}. Reidel. 1981.

Raynal, M. {\em Algorithms for Mutual Exclusion}. MIT Press, 1986. Isbn 
0-262-18119-3.

Robert Rosen, {\em Anticipatory Systems}. Pergamon Press, 1985.

Rosen, R. {\em Life Itself - A Comprehensive Inquiry into the Nature, Origin, 
and Fabrication of Life.} Columbia University Press, 1991. ISBN 
0-231-07564-2.

Roth, J.P. ``An Application of Algebraic Topology to Numerical Analysis: On the 
Existence of a Solution to the Network Problem''. Proceedings of the US 
National Academy of Science, v.45, 1955.

Simon, H. The Sciences of the Artificial. MIT Press 1982 (reprinted from ca. 
1967).

Tonti, E. ``On the formal structure of the relativistic gravitational theory''. 
Accademia Nazionale Dei Lincei, Rendiconti della classe di Scienze fisiche, 
matematiche e naturali. Serie VIII, vol. LVI, fasc. 2 - Feb. 1974. (In 
english.)

www. Various phase web and Topsy publications, including (soon) code 
distribution, are available via {\em www.cs.auc.dk/topsy}.

\end{document}